\newif\ifws
\wsfalse
\ifws

\documentclass[journal,article,accept,oneauthor,pdftex,10pt,a4paper]{mdpi}
\firstpage{1} 
\articlenumber{x}
\doinum{10.3390/------}
\pubvolume{18}
\pubyear{2016}
\copyrightyear{2016}
\externaleditor{Academic Editors: Gregg Jaeger, Andrei Khrennikov and Kevin H. Knuth}
\history{Received: 26 January 2016; Accepted: 19 April 2016; Published: date}

\usepackage{amssymb}
\usepackage[usenames,dvipsnames]{xcolor}
\usepackage{soul}
\usepackage{microtype}

 \theoremstyle{mdpi}
 \newcounter{thm}
 \newcounter{ex}
 \newcounter{re}
 \theoremstyle{mdpidefinition}

\Title{Orthogonal Vector Computations}

\Author{Karl Svozil $^{1,2}$}

\address{%
$^{1}$\quad Institute for Theoretical Physics, Vienna University of Technology,
Wiedner Hauptstrasse 8-10/136, \linebreak Vienna 1040, Austria;  svozil@tuwien.ac.at; Tel.: +43-1-58801  (ext. 13614)
$^{2}$\quad Department of Computer Science, University of Auckland, Private Bag 92019,  Auckland 1142, New Zealand

}

\abstract{Quantum computation is the suitable orthogonal encoding of possibly holistic functional properties into state vectors, followed by a projective measurement.}



\begin{document}

\else

\documentclass[%
 10pt,
 reprint,
  twocolumn,
 showpacs,
 showkeys,
 preprintnumbers,
 amsmath,amssymb,
 aps,
  pra,
  longbibliography,
 ]{revtex4-1}

\usepackage[breaklinks=true,colorlinks=true,anchorcolor=blue,citecolor=blue,filecolor=blue,menucolor=blue,urlcolor=blue,linkcolor=blue]{hyperref}
\usepackage{url}

\usepackage{graphicx}

\usepackage[usenames,dvipsnames]{xcolor}

\begin{document}

\title{Orthogonal vector computations}


\author{Karl Svozil}
\affiliation{Institute for Theoretical Physics, Vienna
    University of Technology, Wiedner Hauptstra\ss e 8-10/136, A-1040
    Vienna, Austria}
\email{svozil@tuwien.ac.at} \homepage[]{http://tph.tuwien.ac.at/~svozil}


\begin{abstract}
Quantum computation is the suitable orthogonal encoding of possibly holistic functional properties into state vectors, followed by a projective measurement.
\end{abstract}

\maketitle
\fi

\section{Identifying Quantum Physical Means for Computation}

The hypothesis pursued in this paper is that
the power of quantum computation solely resides in a proper ``translation'' of ``holistic'' properties of  functions---manifesting themselves in the {\em relational}
 values for different elements of their domain, or of their entire image---into orthogonal subspaces
and their associated perpendicular projections.
This is usually facilitated by {\em {quantum parallelism}}---the possibility to co-represent and co-encode classically distinct and mutually exclusive clauses
into simultaneous coherent superpositions thereof.
However, in order to take advantage of parallelism, it is necessary to be able to analyse the resulting quantum state by suitably chosen
orthogonal projections.
In what follows we shall therefore attempt to enumerate conditions under which a given algorithmic task can be quantum mechanically encoded into
orthogonal subspaces, thereby identifying criteria for potential quantum speedups.

This paper is organized as follows: after a brief and somewhat iconoclastic exposition of the standard quantum formalism
(which may be amusing to some but considered to be superfluous by others)
strategies to encode properties of computable functions into orthogonal subspaces are discussed.
This involves the possibility to orthogonalise non-orthogonal vectors of some initial Hilbert space by interpreting them as
orthogonal projections of mutually orthogonal vectors in a Hilbert space of greater dimension.

\section{Locating Quantum Resources}

Starting with Plack's~\cite{hermann} (p. 31)  ``act of despair''
it may not be totally unreasonable to state that the newly discovered quantum capacities,
unheard of in classical continuum mechanics,
caused shock \linebreak and awe
among some of the most prominent creators of quantum mechanics.
Most notably Schr\"odinger struggled with
quantum coherence, today known as quantum parallelism, throughout his entire life,
{bringing forward seemingly absurd consequences of the formalism, such as the cat paradox,
or quantum} jellification~\cite{schroedinger-interpretation}.
In the same series of papers~\cite{schrodinger} he also mentioned the capacity that
multiple quanta may be entangled;
a resource related to quantum non-locality, which was put forward by Einstein on
various occasions~\cite{einstei-letter-to-schr,Howard1985171}
to question the consistency of quantum and relativity theory.

On the formal side, Hilbert space quantum mechanics~\cite{dirac} and, in particular,
quantum logic have been conceived~\cite{birkhoff-36} as attempts
to systematically identify and study the valid, feasible, operational resources and capacities
of quantized physical systems in terms of algebraic structures arising in {quantum mechanics.
Thereby, logical entities are identified with quantum theoretical terms;
\linebreak a typical }example being the identification with a binary logical yes-no proposition or property
with an orthogonal projection operator;
or, equivalently, with the associated linear subspace, which, in the one-dimensional case,
is spanned by a single (unit) vector.
In what follows some of these identifications will explicitly be enumerated.

Subsequently, and possibly motivated by single-quantum experiments which have become experimentally feasible,
quantum information and communication research attempts to exploit mostly non-classical capacities of
quantized systems, such as parallelism, entanglement, or other quantum resources~\cite{be-va,benn:97,fortnov-03,Castagnoli}.
(A {\it{caveat:}} we refrain here from a detailed, more comprehensive list of references,
as~this might transcend this brief review.)

\subsection{State as Context}

In the Dirac-von Neumann formalism of quantum mechanics a {\em{pure state}}  is completely
characterized by a unit vector
$\vert \psi \rangle$ (or, equivalently, by the associated one-dimensional subspace
spanned by $\vert \psi \rangle$;
or by the orthogonal projection operator $\textsf{\textbf{E}}_\psi = \vert \psi \rangle \langle \psi \vert$)
in a Hilbert space, which~essentially
(disregarding completeness) is
a vector space equipped with a scalar product.
In what follows we shall disregard mixed states completely---according to the {\em {Church of the Larger Hilbert Space}} any mixed state is purely epistemic (but not ontic),
and originates from some unknown pure state which can be reconstructed, subject to some (usually non-unique) purification
by ``reverting'' (by disclosing ``additional information'') the introduction of ignorance
(enforced by a non-unitary many-to-one state evolution)
through partial traces.
Every state will thus assumed to be pure.

Alternatively it may be preferable~\cite{svozil-2013-omelette} to define a state { via} a context, or, equivalently,  an~orthonormal basis,
or a maximal observable~\cite{halmos-vs} (\S~84, Theorem~1) whose spectral sum is~non-degenerate.
Any such context represents a Boolean subalgebra (formed by mutually orthogonal and thus commuting projections) of a quantum logic;
a kind of ``greatest classical mini-universe'' which is co-prepareable.

\subsection{Observable as Context}

Every maximal {\em {observable}} or context corresponds to the non-degenerate spectral sum of one-dimensional mutually orthogonal (and thus commuting)
projections whose sum is a resolution of the identity.
Within operational capacities, those bases can be chosen freely by the experimenter.\linebreak
They correspond to quasi-classical mini-universes of mutually commuting, compatible observables.

Once a particular maximal observable is chosen, the respective orthonormal basis\linebreak
$\mathfrak{B}_n =\left\{ \vert {\bf e}_1 \rangle, \vert {\bf e}_2 \rangle, \ldots , \vert {\bf e}_n \rangle
\right\}$ facilitates a {\em {particular view}} on the state $\vert \psi \rangle$.
If one basis element coincides with the state $\vert \psi \rangle$,
then the measurement of $\mathfrak{B}_n$ is deterministic:
a single proposition associated with the orthogonal projection operator identical to
$\vert \psi \rangle \langle \psi \vert$
is true; and all the other propositions represented by the remaining basis elements are false.
In this case we may say that this observable
reflects a physical {\em {property}} of that state~\cite{2015-AnalyticKS}
(thereby affirming the dual use of orthogonal projections both for pure states as well as for propositions).

Otherwise, if there is a mismatch between the state preparation and the query, {\em {context translation}} introduces stochasticity
through the many degrees of freedom of quasi-classical measurement devices~\cite{svozil-2003-garda}.
The epistemic interpretation of maximal observables as views on the state $\vert \psi \rangle$
alleviates
Schr\"odinger's continued concerns regarding the ontological existence of coherent superpositions,
so~vividly put forward in the cat paradox or by quantum jellification
mentioned in his lesser known
Dublin seminars \cite{schroedinger-interpretation}.

\subsection{Probability}

Gleason's theorem~\cite{Gleason} states that (for dimensions larger than two, allowing intertwining orthogonal bases) the quantum probabilities
follow from the requirement that on all contexts the quantum probabilities should be identical with their respective classical probabilities
(in particular, $\sigma$-additivity of probabilities of mutually exclusive events). That~is, within classical mini-universes of a quantum logic,
classical probabilities rule.

Although most of his paper is devoted to proving necessity of the Born rule (including mixed states) from this assumption,
already in the second paragraph Gleason observed that,
as long as one is concerned merely about pure states
it is possible to construct a {\em {probability measure}} on the Hilbert space
{via} an elementary geometric construction, suggesting the squared norm of the orthogonal projections
of these pure states onto the vectors of some orthonormal basis
as quantum probability measure.
One could again say that the pure state is ``viewed from'' the maximal operator corresponding to the aforementioned basis.

In contrast to the full proof of Gleason's theorem,
the argument for the sufficiency of the squared norm  probability measures (satisfying classicality within contexts)
is elementary;
thereby the absolute squares of the probability amplitudes
$\langle  \psi \vert {\bf e}_i  \rangle$ result from the Pythagorean theorem:
since $\vert \psi \rangle$ is a unit vector and
the  $\vert {\bf e}_i  \rangle$ represent vectors of an orthonormal basis,
all absolute squares of $\langle  \psi \vert {\bf e}_i  \rangle$  sum up to~one;
that is,
$\sum_{i=1}^n \vert\langle  \psi \vert {\bf e}_i  \rangle \vert^2=1$.
A construction of such a Gleason-type measure
in two-dimensional real vector space is depicted in Figure~\ref{2015-m-fdlvs-vv}.
\begin{figure}
\begin{center}
\unitlength 0.35mm
\linethickness{0.4pt}
\ifx\plotpoint\undefined\newsavebox{\plotpoint}\fi 
\begin{picture}(178.25,188)(25,30)
\multiput(167.18,109.93)(0,-2){15}{\color{Red}\line(0,-1){0.5}}
\multiput(166.93,109.93)(-2,0){48}{\color{Red}\line(-1,0){0.5}}
\multiput(166.93,109.93)(-2,-2){32}{\color{ForestGreen}\line(-1,0){0.5}}
\multiput(166.93,109.93)(-2,2){17}{\color{ForestGreen}\line(-1,0){0.5}}
%
%
\put(71.5,81.25){\color{Red}\vector(1,0){110}}
\put(71.5,81.25){\color{Red}\vector(0,1){100}}
\put(167,109.75){\color{blue}\vector(3,1){.07}}\multiput(71.5,81.25)(.11301775148,.03372781065){845}{\color{blue}\line(1,0){.11301775148}}
\put(71.5,81.25){\color{ForestGreen}\vector(1,1){70}}
\put(71.5,81.25){\color{ForestGreen}\vector(-1,1){50}}
\put(71.5,81.25){\color{ForestGreen}\vector(1,-1){50}}
\put(175,111.25){\makebox(0,0)[cc]{\color{blue}\scriptsize $\vert \psi  \rangle$}}
\put(195,81.5){\makebox(0,0)[cc]{\color{Red}\scriptsize $\vert {\bf e}_1 \rangle$}}
\put(147.75,157){\makebox(0,0)[cc]{\scriptsize $\vert {\bf f}_1 \rangle$}}
\put(71.5,188){\makebox(0,0)[cc]{\color{Red}\scriptsize $\vert {\bf e}_2 \rangle$}}
\put(20,140){\makebox(0,0)[cc]{\color{ForestGreen}\scriptsize $\vert {\bf f}_2 \rangle$}}
\put(125,25){\makebox(0,0)[cc]{\color{ForestGreen}\scriptsize $- \vert {\bf f}_2 \rangle$}}
\put(135,120){\makebox(0,0)[cc]
{\color{Red}\scriptsize $\vert \langle \psi  \vert {\bf e}_1 \rangle \vert$}}
\put(185,97){\makebox(0,0)[cc]
{\color{Red}\scriptsize $\vert \langle \psi  \vert {\bf e}_2 \rangle \vert$}}
\put(145,64.5){\makebox(0,0)[cc]
{\color{ForestGreen}\scriptsize $\vert \langle \psi  \vert {\bf f}_1 \rangle \vert$}}
\put(170,135){\makebox(0,0)[cc]
{\color{ForestGreen}\scriptsize $\vert \langle \psi  \vert {\bf f}_2 \rangle \vert$}}
\end{picture}
\end{center}
\caption{Different orthonormal bases
{\color{black}
$\{
\vert {\bf e}_1 \rangle ,
\vert {\bf e}_2 \rangle
\}$}
and
{\color{black}
$\{
\vert {\bf f}_1 \rangle ,
\vert {\bf f}_2 \rangle
\}$}
offer different ``views''
on the pure state {\color{black} $\vert \psi  \rangle$}.
As {\color{black} $\vert \psi  \rangle$} is a unit vector
it follows  from the Pythagorean theorem that
${\color{black}
\vert \langle \psi  \vert {\bf e}_1 \rangle \vert^2
+
\vert \langle \psi  \vert {\bf e}_2 \rangle \vert^2}=
{\color{black}
\vert \langle \psi  \vert {\bf f}_1 \rangle \vert^2
+
\vert \langle \psi  \vert {\bf f}_2 \rangle \vert^2}
=1
$, thereby
motivating the use of the abolute value (modulus) squared of the amplitude for
quantum probabilities on pure states.}
  \label{2015-m-fdlvs-vv}
\end{figure}

By construction, this probability measure satisfies
the requirement of the validity of classical probabilities within all classical mini-universes.
This is remarkable because of the absence of (sufficiently many) global two-valued measures~\cite{kochen1}
whose positive convex combinations form all classical probability measures~\cite{pitowsky-89a}.

\subsection{Entanglement}

Pointedly stated, quantum entanglement is a consequence of the inseparability of not-so-individual quanta.
In general, quantum states do not allow the decomposition into
product states of constituents.
For instance, a general three partite state of three bits
$\vert \psi \rangle =\sum_{i,j,k=0}^1 \alpha_{ijk} \vert ijk\rangle$
is entangled if and only if it cannot be written as a single product state of the three constituents;\linebreak
a non-entangled product state of three particles requires~\cite{mermin-07} (p.~18) that
all of the following equations are satisfied:
$\alpha_{000} \alpha_{011} = \alpha_{001} \alpha_{010}$,
$\alpha_{000} \alpha_{101} = \alpha_{001} \alpha_{100}$,
$\alpha_{000} \alpha_{110} = \alpha_{010} \alpha_{100}$,
$\alpha_{000} \alpha_{111} = \alpha_{011} \alpha_{100}$,
$\alpha_{001} \alpha_{110} = \alpha_{011} \alpha_{100}$,
$\alpha_{001} \alpha_{111} = \alpha_{011} \alpha_{101}$,
$\alpha_{010} \alpha_{101} = \alpha_{011} \alpha_{100}$,
$\alpha_{010} \alpha_{111} = \alpha_{011} \alpha_{110}$,
$\alpha_{100} \alpha_{111} = \alpha_{101} \alpha_{110}$.

As a consequence, unlike classical properties, entangled states
encode (not purely individual but) {\em {relational}} properties of multiple quanta~\cite{Zeilinger-97,zeil-99}.
It is thus futile to request, as Einstein maintained in a~letter communicated to Schr\"odinger~\cite{einstei-letter-to-schr,Einstein-48,Howard1985171}, that
the sub-state of an entangled state is not affected by any change of another sub-state if the two involved entangled sub-systems are
spatially separated.

\subsection{Evolution as Permutation}

The quantum evolution is postulated to be an isometry (that is, a distance preserving map);\linebreak
or, stated differently, a permutation---a one-to-one mapping---preserving the scalar product.
This~can be formalized by unitary transformations~\cite{Schwinger.60}:
suppose
$\mathfrak{B}_n= \{\vert {\bf e}_1\rangle , \vert  {\bf e}_2\rangle , \ldots , \vert {\bf e}_n\rangle \}$
and\linebreak
$\mathfrak{B}'_n= \{\vert {\bf f}_1\rangle , \vert  {\bf f}_2\rangle , \ldots , \vert {\bf f}_n\rangle \}$
are orthonormal bases, then $
\textsf{\textbf{U}}_{fe}=  \sum_{i=1}^n  \vert {\bf f}_i\rangle \langle {\bf e}_i \vert
$
yields a unitary transformation; conversely, any unitary transformation can be represented by such a change of orthonormal bases.
So, stated pointedly, the quantum evolution amounts to a (generalized) rotation either of the state vector (the active, Schr\"odinger representation),
or of the coordinate frame (passive, Heisenberg picture) relative to which the state is viewed.

Much of the conceptual progress in the foundations of quantum mechanics
owes to the exploitation of this one-to-one permutation scheme in various scenarios:
to name just three examples, take quantum teleportation~\cite{BBCJPW},
induced coherence~\cite{zou-wang-mandel:91a},
and the quantum erasure~\cite{PhysRevD.22.879,PhysRevA.25.2208,engrt-sg-I}.
Indeed,
as~has already been pointed out by Everett~\cite{everett} (p.~454) (see also Wigner~\cite{wigner:mb})
there is no such thing as an
``irreversible measurement'' if quantum theory is assumed universally and uniformly valid:
what~we fapp~\cite{bell-a} call ``measurement'' is means relative~\cite{Myrvold2011237}, epistemic,
and could, at least in principle, be undone if our operational capacities permit.

\subsection{Computational Resources}

Quantum theory offers or identifies a single state vector, or orthonormal systems of vectors, which~can be rotated or viewed from various
perspectives,
and on which projection measurements can be made,
as the quantum resource for computations.
Related alleged capacities, such as  ``quantum parallelism'' associated with coherent superposition induced by views on this vector which do not coincide with it,
or {\it {ex nihilo}}
randomness induced by the quasi-stochasticity of the resulting measurement results through context translations,
or entanglement originating from the pretension that the state of subsystems which are spatially apart must be separable,
are implications
which need to be carefully revised~\cite{jaynes-89,jaynes-90}
in order to re-evaluate the quantum capacities for computation.

\section{Renditions via Orthogonal Subspaces}

So far the stage has been set for concrete proposals to utilize quantized systems for ``optimized'' computation:
in particular, by ``looking at a pure state from a proper perspective;''
that is, \linebreak from an orthonormal basis of which it is not an element.
In such a basis, the pure state has a~ composition in terms of more than one basis vectors---in quantum mechanical terms, the state is in a {\em {coherent superposition}} of these basis vectors.
If the basis ist associated with classical bit states, such as  $\mathfrak{B}_2=\{\vert 0 \rangle ,\vert 1\rangle \}$,
one could claim
that a quantum state such as
$\alpha_0 \vert 0 \rangle + \alpha_1\vert 1\rangle$ with non-vanishing $\alpha_0$, $\alpha_1$
is ``in~both states $\vert 0 \rangle$ and $\vert 1\rangle$ simultaneously.''
This may nurture hopes that the simultaneous co-processing of an arbitrary amount of data or clauses
might become quantum feasible
by introducing a ``sufficient'' amount of bits in superposition.

One fundamental issue which has to be accounted for is the fact that
any recursive function is completely characterized by its input-output behaviour.
In order to perform an evaluation (of the recursive function), this behaviour must be coded into a physical operation {\it {simulacron}}~\cite{simula}.

However, there is a fundamental limit to quantum computation
which resides in one of the basic relations between vectors in Hilbert space: their (non-)orthogonality.
This translates into the criterion that a query is quantum mechanically feasible if and only if it can be encoded into orthogonal~subspaces.
As the unitary evolution preserves angles, and thus orthogonality,
within the same Hilbert space there is nothing one can do about ``orthogonally separating'' two non-orthogonal subspaces,
\linebreak because a Gram-Schmidt orthogonalisation is quantum infeasible
when it is most needed; that is, if the initial vectors are not orthogonal
(just as the cloning of an arbitrary quantum state is impossible for the same reason---unitarity of the quantum state evolution).

\subsection{Example}

Recall
the supposition (or rather speculation) mentioned earlier:
that the result of any kind of (binary) decision problem can be obtained by taking ``enough'' qubits to ``cover all cases''
through the coherent superposition of all classical cases by a single quantum query.
Indeed, for the sake of temporal speedups, we could be so bold as to attempt to quantum encode an algorithm as the ``parallel''
(weighted by relative phases and amplitudes) sum (not the product); that is,  the coherent superposition, of all conceivable classical clauses.
(Never mind computation space; that is, the amount of qubits, as~long as this resource is bounded.
Memory space is also required in order to render the computation one-one; that is, a permutation,
if the evaluated functions are not invertible.)

For the sake of an example of this strategy, consider Deutsch's problem for computing the parity of a binary function of a single bit.
there, two bits---the input bit and an auxiliary bit (to take care of injectivity in case the function is constant)---which are initially classical,
are spread out by Hadamard gates into a proper (the relative phases are relevant) coherent superposition of all classical states.

At the heart of this algorithm is a way to subdivide the four-dimensional state space of the two~bits
into two orthogonal subspaces associated with parity $\pm 1$, respectively,
thereby rendering parity in a single query. (See ~\cite{Farhi-98} for a different finding.)
The quantum {\it versus} classical trade-off is the ignorance of the precise function (out of four) after this single quantum query.

How can this exactly be achieved?
Suppose we identify with the classical bit states $0$ and $1$ the quantum states, represented by the two vectors of a Cartesian basis $\mathfrak{B}_2$,
$0 \equiv \vert 0 \rangle \equiv \begin{pmatrix} 1\\0\end{pmatrix}$
and
$1 \equiv \vert 1 \rangle \equiv \begin{pmatrix} 0\\1\end{pmatrix}$,
respectively.
Suppose further that the four binary functions of a single classical bit are  the two constant  functions
$f_0(x)=0$ and
$f_3(x)=1$,
as well as the identity $f_1(x) = x$ and the negation $f_2(x) = 1 \oplus x$, where $x \in \{0,1\}$ and $\oplus$ represents the addition modulo $2$.
By forming a unitary quantum oracle $\textsf{\textbf{U}}_{f_i} \vert {\bf x} {\bf y} \rangle = \vert {\bf x} \left[{\bf y}\oplus f_i({\bf x})\right] \rangle$, $i\in \{0,1,2,3\}$,
and by taking $\vert {\bf x}  \rangle = \vert {\bf y}  \rangle = \left(1/\sqrt{2}\right)\vert 0-1 \rangle $
one obtains   (by omitting normalization factors; the superscript $^T$ indicates transposition, and ``$\pm$'' stands for ``$\pm 1$,'' respectively)
\setlength{\medmuskip}{1mu}
\begin{equation}
\begin{split}
\textsf{\textbf{U}}_{f_0}\vert (0-1)(0-1) \rangle = \vert +00 - 01 - 10 + 11 \rangle \equiv  \begin{pmatrix} +--+\end{pmatrix}^T   \\
\textsf{\textbf{U}}_{f_1}\vert (0-1)(0-1) \rangle = \vert +00 - 01 + 10 - 11 \rangle \equiv  \begin{pmatrix} +-+-\end{pmatrix}^T  \\
\textsf{\textbf{U}}_{f_2}\vert (0-1)(0-1) \rangle = \vert -00 + 01 - 10 + 11 \rangle \equiv  \begin{pmatrix} -+-+\end{pmatrix}^T \\
\textsf{\textbf{U}}_{f_3}\vert (0-1)(0-1) \rangle = \vert -00 + 01 + 10 - 11 \rangle \equiv  \begin{pmatrix} -++-\end{pmatrix}^T
\end{split}
\label{2016-vector-e-dp}
\end{equation}
\setlength{\medmuskip}{3mu}

If one is being presented with an unknown function $f_i$
and is given the task to identify~\cite{e-f-moore,svozil-2001-eua} it,
the following two-(binary)-valued query is associated with functional parity:
in terms of subspaces in fourdimensional Hilbert space
with the basis
$\mathfrak{B}_2 \equiv \{\vert {\bf e}_1\rangle , \vert  {\bf e}_2\rangle \}
\equiv
\left\{
(1/2) \begin{pmatrix} 1,-1,-1,1\end{pmatrix}^T ,
(1/2) \begin{pmatrix} 1,-1,1,-1\end{pmatrix}^T
\right\}
$, and written as spectral sum
$
\textsf{\textbf{P}} =
 \vert {\bf e}_1 \rangle \langle {\bf e}_1 \vert
-
 \vert {\bf e}_2 \rangle \langle {\bf e}_2 \vert
$
corresponding to a partitioning~\cite{DonSvo01,svozil-2002-statepart-prl,2007-tkadlec-svozil-springer}
of functions
$
\{\{f_0,f_3\},\{f_1,f_2\}\}
$.

It might not be too critical to call the coincidence
between the particular ``modulo two'' oracle and parity of a binary function of one bit
{\it ad hoc.}

\subsection{Generalized Gram-Schmidt Process}

In the previous example the input-output behaviour
$\{\{0,0\},\{0,1\},\{1,0\},\{1,1\}\}$
of a functional class
has been translated into the appropriate orthogonal subspaces;
the trade-off being the enlargement of Hilbert space by two extra dimensions.
The question arises if it is always possible to do so.

\begin{figure}
\begin{center}
\unitlength 0.5mm 
\linethickness{0.4pt}
\ifx\plotpoint\undefined\newsavebox{\plotpoint}\fi 
\begin{picture}(118,100)(0,25)
\thinlines
\put(50,50){\color{Red}\vector(1,-1){25}}
\put(82,25){\makebox(0,0)[cc]{\color{Red}\scriptsize $\vert {\bf f}_2 \rangle$}}
\thicklines
\put(75,62.25){\color{Red}\vector(2,1){.07}}\multiput(50,50)(.0686813187,.0336538462){364}{\color{Red}\line(1,0){.0686813187}}
\put(82,65){\makebox(0,0)[cc]{\color{Red}\scriptsize $\vert {\bf e}_2 \rangle$}}
\thinlines
\put(50,50){\color{ForestGreen}\vector(1,1){50}}
\put(107,100){\makebox(0,0)[cc]{\color{ForestGreen}\scriptsize $\vert {\bf f}_1 \rangle$}}
\thicklines
\put(50,50){\color{ForestGreen}\vector(1,0){50}}
\put(106,50){\makebox(0,0)[cc]{\color{ForestGreen}\scriptsize $\vert {\bf e}_1 \rangle$}}
\thinlines
\multiput(99.93,49.93)(0,.980392){52}{{\color{ForestGreen}\rule{.4pt}{.4pt}}}
\multiput(75,62.156)(0,-.997144){38}{{\color{Red}\rule{.4pt}{.4pt}}}
\multiput(51.25,35)(.2367021277,.0336879433){282}{\line(1,0){.2367021277}}
\multiput(11.25,75)(.2367021277,.0336879433){282}{\line(1,0){.2367021277}}
\put(118,44.5){\line(-1,1){40}}
\put(50.75,35){\line(-1,1){40}}
\qbezier(73.158,60.755)(81.094,57.707)(78.729,50.033)
\put(82.618,56.235){\makebox(0,0)[lc]{$\varphi$}}
\put(62,44.8){\makebox(0,0)[lc]{$\frac{\pi}{2}$}}
\thinlines
\multiput(56.041,55.893)(.034149529,-.03364733){148}{\line(1,0){.034149529}}
\multiput(61.096,50.913)(-.033593421,-.034013339){177}{\line(0,-1){.034013339}}
\end{picture}
\end{center}
\caption{Demonstration of the orthogonalisation by dimensional lifting for two planar vectors
{\color{black}$\vert {\bf e}_1 \rangle$}
and {\color{black} $\vert {\bf e}_2 \rangle$}
which are ``lifted'' into two orthogonal vectors
{\color{black} $\vert {\bf f}_1 \rangle$}
and
{\color{black} $\vert {\bf f}_2 \rangle$} in three dimensions which orthogonally project
onto {\color{black}$\vert {\bf e}_1 \rangle$}
and {\color{black} $\vert {\bf e}_2 \rangle$}.
}
  \label{2016-vector-f2}
\end{figure}
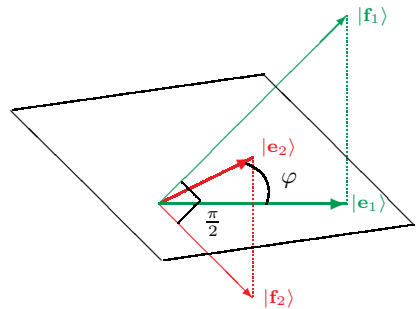

As an illustration consider two non-orthogonal and non-collinear planar vectors depicted in Figure~\ref{2016-vector-f2}.
Suppose we ``extend'' those vectors into the three-dimensional space such that the
direction and length of the two new non-planar vectors is chosen such that
(i) they are orthogonal,
and at the same time
(ii) their orthogonal projections onto the plain
are identical with the original planar vectors.

In the following we shall prove the formal (but not quantum physical)
possibility of {\em {orthogonalisation by dimensional lifting:}}
An arbitrary number $k$ of nonzero vectors
$\{ \vert {\bf e}_1 \rangle ,\ldots ,\vert {\bf e}_k \rangle \}$
of $n$-dimensional Hilbert space can be (non-uniquely)
interpreted as the orthogonal projections (onto the original Hilbert space)
of a set of mutually orthogonal
(\emph{i.e.}, $\langle {\bf f}_i \vert {\bf f}_j \rangle =0$ for $i\neq j$)
vectors  $\{\vert {\bf f}_1 \rangle ,\ldots \vert {\bf f}_k \rangle  \}$
in an Hilbert space of dimension $m\ge n$.

For the sake of an explicit example, suppose that, for two vectors (\emph{i.e.}, $k=2$) and arbitrary finite dimension $n$,
we start with
$\vert {\bf e}_1 \rangle \equiv \begin{pmatrix}x_{11},\ldots ,x_{1n}\end{pmatrix}^T$,
and
$\vert {\bf e}_2 \rangle \equiv \begin{pmatrix}x_{21},\ldots ,x_{2n}\end{pmatrix}^T$.
It is not too difficult to construct two vectors
\begin{equation}
\begin{split}
\vert {\bf f}_1 \rangle \equiv \begin{pmatrix}x_{11},\ldots ,x_{1n},1\end{pmatrix}^T   \\
\vert {\bf f}_2 \rangle \equiv \begin{pmatrix}x_{21},\ldots ,x_{2n},- \left[x_{11} x_{21} + \cdots  + x_{1n} x_{2n}\right]\end{pmatrix}^T
\end{split}
\end{equation}
which are orthogonal vectors in $n+1$ dimensions and project onto
$\vert {\bf e}_1 \rangle$ and
$\vert {\bf e}_2 \rangle$,
respectively.
The~configuration for $n=2$ is depicted in Figure~\ref{2016-vector-f2}.

The general proof by construction is iterative and follows an ``inverse'' (in the sense of ``adding'' {some dimension rather than ``subtracting'' existing vector projections)
generalized Gram-Schmidt process:}
suppose all the $k$ vectors of $n$-dimensional space are non-collinear and non-orthogonal.
\linebreak Then,  by a series of  combinations, whose number is determined by the binomial coefficient $\begin{pmatrix} k \\ 2 \end{pmatrix}$, \linebreak of two vector orthogonalisations as above,
while at the same time, by taking the new vector component of all the other $k-2$ vectors zero,
one can construct a set of $k$ mutually orthogonal vectors in
$\left[k + \begin{pmatrix} k \\ 2 \end{pmatrix}\right]$-dimensional vector space whose orthogonal projections on the original space are the original vectors.
Another orthogonalisation by dimensional lifting requiring merely $n+k$ dimensions due to Havlicek~\cite{havlicek-priv2}
will be discussed in a forthcoming paper.

A necessary and sufficient condition for a generalized Gram-Schmidt process to be representable by a quantum evolution is
its  correspondence to some unitary operation from the initial into \linebreak the final state.
Unfortunately, by design this generalized Gram-Schmidt process is not unitary {\it {per se}}.
(However, this also is true for the functional encoding of bits before the introduction of \linebreak a quantum oracle:
{\it a priori} the introduction of auxiliary bits is not guaranteed to be representable by a unitary transformation.)
Recall that, as has been mentioned earlier, a necessary and sufficient condition for a unitary transformation is that, by
$
\textsf{\textbf{U}}_{fe}=  \sum_{i=1}^n  \vert {\bf f}_i\rangle \langle {\bf e}_i \vert
$
any orthonormal basis
$\mathfrak{B}_n \equiv \{\vert {\bf e}_1\rangle , \vert  {\bf e}_2\rangle , \ldots , \vert {\bf e}_n\rangle \}$ is transformed
into another one
$\mathfrak{B}'_n \equiv \{\vert {\bf f}_1\rangle , \vert  {\bf f}_2\rangle , \ldots , \vert {\bf f}_n\rangle \}$~\cite{Schwinger.60}.
However, in our case, neither are the original vectors orthogonal, nor needs the target system to be a basis.\linebreak
(This latter deficiency could, in principle, be remedied by adding basis vectors,
thereby creating an~eutactic star~\cite{schlaefli-1901,hasse-stachel96}.)
So, unless the vectors resulting from (i) are not already properly orthogonal (for the particular quantum query),
the generalized Gram-Schmidt process (ii) has no quantum mechanical realization.

Moreover, unitary transformations preserve the angles among vectors.
Therefore, if the functional behaviour, subjected to a quantum oracle,
cannot separate different functional properties by orthogonal subspaces,
not much can be gained by a straightforward generalized Gram-Schmidt processes.

Nevertheless, let us, for the sake of another example of dimensional lifting,
come back to \linebreak Deutsch's problem, and
consider another {\it ad hoc} quantum oracle, namely
$\textsf{\textbf{V}}_{f_i} \vert {\bf x} {\bf y} \rangle = \vert {\bf x}  f_i({\bf y}) \rangle$, $i\in \{0,1,2,3\}$,
resulting in
\setlength{\medmuskip}{1mu}
\begin{equation}
\begin{split}
\textsf{\textbf{V}}_{f_0}\vert (0+1)(0+1) \rangle = \vert +00 + 10 \rangle \equiv  \begin{pmatrix} +0+0\end{pmatrix}^T    \\
\textsf{\textbf{V}}_{f_1}\vert (0+1)(0+1) \rangle = \vert +00 + 11 \rangle \equiv  \begin{pmatrix} +00+\end{pmatrix}^T  \\
\textsf{\textbf{V}}_{f_2}\vert (0+1)(0+1) \rangle = \vert +01 + 10 \rangle \equiv  \begin{pmatrix} 0++0\end{pmatrix}^T   \\
\textsf{\textbf{V}}_{f_3}\vert (0+1)(0+1) \rangle = \vert +01 + 11 \rangle \equiv  \begin{pmatrix} 0+0+\end{pmatrix}^T
\end{split}
\label{2016-vector-e-dv}
\end{equation}
\setlength{\medmuskip}{3mu}

By appending four extra dimensions it is not difficult to {\it ad hoc} orthogonalise this set of vectors (two pairs are already orthogonal);
namely
\begin{equation}
\begin{split}
\begin{pmatrix} +0+0+000\end{pmatrix}^T   \\
\begin{pmatrix} +00+-+00\end{pmatrix}^T  \\
\begin{pmatrix} 0++0--++\end{pmatrix}^T  \\
\begin{pmatrix} 0+0+0---\end{pmatrix}^T
\end{split}
\label{2016-vector-e-dvo}
\end{equation}
so that, provided a suitable quantum oracle could be found which maps the functional behaviour into
the orthogonal subspaces spanned by the vectors in {Equation} (\ref{2016-vector-e-dvo}),
an unknown function $f_i$ could be identified uniquely
by a single query (encoding the orthogonal projections corresponding to this latter set of orthogonal vectors).

\subsection{Counterexample and (In-)Sufficiency}

Our hypothesis has been that it is not totally unreasonable
to speculate that any functional behaviour can somehow be mapped into orthogonal subspaces
through the application of a single functional call whose argument is the (equal-weighted) superposition of all possible classical variations of arguments.
More specifically, if reversibility is guaranteed by the introduction of auxiliary~bits,
{different functional behaviours on such arguments result in mutually different (non-orthogonal) vectors;} which then might be
``lifted''  into mutually orthogonal subspaces in Hilbert spaces of greater dimensions than the original Hilbert space.
Any query, performed on this higher-dimensional {Hilbert~space, could be realized by ``bundling''
(formally by adding)
the associated projections appropriately.}

Indeed, it might be possible (in contrast to findings using different assumptions and oracles~\cite{Farhi-98})
with this method to solve the parity problem by a single quantum query, through separating the parity even and odd functions into
two orthogonal subspaces, whose direct sum is the entire Hilbert space.

However, the unitary quantum state evolution imposes strong restrictions on such queries, because it essentially amounts to isometries; that is,
to rotations of state vectors:
the functional properties must be encoded into these rotations; and
the subspaces associated with the particular functional classes
which need to be distinguished from one another must be
rotated in such ways that their relative separation amounts to orthogonality.


Therefore there exist collections of functions which, relative to some particular quantum oracles,
in particular,
$\textsf{\textbf{U}}_i( \vert {\bf x}_1,\ldots , {\bf x}_n, {\bf y} \rangle ) =
\vert {\bf x}_1,\ldots , {\bf x}_n, {\bf y}\oplus f_i(   {\bf x}_1,\ldots , {\bf x}_n)\rangle$,
cannot be orthogonally separated.

For an indication consider~\cite{svozil-2015-AUC-svozil}, as a generalization of
Deutsch's problem discussed earlier, two~particular binary functions of two bits with different parity,
namely
$f_0(xy)=0$ for all $x,y\in \{0,1\}$,
and
$f_{8}(00)=f_{8}(01)=f_{8}(10)=0$, as well as $f_{8}(11)=1$.
A similar calculation as in Equation ~(\ref{2016-vector-e-dp})~yields
\begin{equation}
\begin{split}
\textsf{\textbf{U}}_{f_0}\vert (0-1)(0-1)(0-1) \rangle \\
= \vert + 000 - 001 - 010 + 011 - 100 + 101 + 110 - 111 \rangle \\
\equiv  \begin{pmatrix} +--+-++-\end{pmatrix}^T    \\
\textsf{\textbf{U}}_{f_8}\vert (0-1)(0-1)(0-1) \rangle \\
= \vert + 000 - 001 - 010 + 011 - 100 + 101 - 110 + 111 \rangle \\
\equiv  \begin{pmatrix} +--+-+-+\end{pmatrix}^T
\end{split}
\label{2016-vector-e-dpgen}
\end{equation}
$f_0$ and $f_8$ represent functional parity $+1$ and $-1$; and yet
their respective quantum oracles  $\textsf{\textbf{U}}_{f_0}$ and $\textsf{\textbf{U}}_{f_8}$
do not map the state $\vert (0-1)(0-1)(0-1) \rangle$ into orthogonal states
(indeed, non-orthogonality may come as no surprise).

Note that, although the relative phases of the coherent superposition of the three input bits have been chosen to be $\pi$,
other phases could result in orthogonal vectors as desired.
For the sake of demonstration, consider the general case (with $\vert a_i \vert^2  + \vert b_i \vert^2 = 1$)
\begin{widetext}
\begin{equation}
\begin{split}
\textsf{\textbf{U}}_{f_0}
\left( a_1\vert 0 \rangle + b_1\vert 1 \rangle \right)
\left( a_2\vert 0 \rangle + b_2\vert 1 \rangle \right)
\left( a_3\vert 0 \rangle + b_3\vert 1 \rangle \right)
\\
=
a_1a_2a_3 \vert 000 \rangle  +
a_1a_2b_3 \vert 001 \rangle  +
a_1b_2a_3 \vert 010 \rangle  +
a_1b_2b_3 \vert 011 \rangle  + \\
b_1a_2a_3 \vert 100 \rangle  +
b_1a_2b_3 \vert 101 \rangle  +
b_1b_2a_3 \vert 110 \rangle  +
b_1b_2b_3 \vert 111 \rangle \\
\equiv
\begin{pmatrix}
a_1a_2a_3 ,
a_1a_2b_3 ,
a_1b_2a_3 ,
a_1b_2b_3 ,
b_1a_2a_3 ,
b_1a_2b_3 ,
b_1b_2a_3 ,
b_1b_2b_3
\end{pmatrix}^T  \\
\textsf{\textbf{U}}_{f_8}
\left( a_1\vert 0 \rangle + b_1\vert 1 \rangle \right)
\left( a_2\vert 0 \rangle + b_2\vert 1 \rangle \right)
\left( a_3\vert 0 \rangle + b_3\vert 1 \rangle \right)
\\
=
a_1a_2a_3 \vert 000 \rangle  +
a_1a_2b_3 \vert 001 \rangle  +
a_1b_2a_3 \vert 010 \rangle  +
a_1b_2b_3 \vert 011 \rangle  +  \\
b_1a_2a_3 \vert 100 \rangle  +
b_1a_2b_3 \vert 101 \rangle  +
b_1b_2b_3 \vert 110 \rangle  +
b_1b_2a_3 \vert 111 \rangle \\
\equiv
\begin{pmatrix}
a_1a_2a_3 ,
a_1a_2b_3 ,
a_1b_2a_3 ,
a_1b_2b_3 ,
b_1a_2a_3 ,
b_1a_2b_3 ,
b_1b_2b_3 ,
b_1b_2a_3
\end{pmatrix}^T
\end{split}
\label{2016-vector-e-dpgen1}
\end{equation}
\end{widetext}

The scalar product of these vector vanishes for the particular value assignments
 $a_1=a_2=a_3=0$ and thus $b_1=b_2=b_3=1$.
Alas, this particular value assignment is improper to separate other cases
such as, for instance, $f_8$ and $f_{15} = 1$
orthogonally.

\section{Summary}

The main goal of this paper has been a discussion of, and the enumeration of criteria for encoding some algorithmic task into quantum mechanically feasible orthogonal subspaces.
This is by no means a trivial task.

With regards to somehow ``mapping'' a functional behaviour into Hilbert space,
such that certain  relational or  holistic functional properties---manifesting themselves  in the {\em {relational}} values for different elements of their domain,
and only after the evaluation of more than one classical input---reveal themselves through efficient quantum queries,
the situation is what may be called ``ambivalent.''

On the one hand, relative to some quantum oracle, it is sometimes impossible to decide certain
relational or functional properties,
simply because this particular quantum oracle is unsuitable to render the appropriate orthogonal subspaces.

On the other, more promising, hand, by the orthogonalisation via dimensional lifting of non-orthogonal vectors,
it cannot be excluded that any functional property can be mapped into orthogonal subspaces of some Hilbert space of higher dimension
than the space spanned by the original non-orthogonal vectors.
In this sense, any functional property, including parity, might be decidable by a single quantum query.

The situation is not dissimilar to the purification of a mixed state, where the
``missing information'' is (non-uniquely) supplemented by fitting hypotheses or conjectures (about the original pure state).
However, this generalized Gram-Schmidt process is not unitary in general, and thus does not correspond to any quantum capacity.

So far, all quantum speedups based on orthogonalisation have been {\it ad hoc};
no method to convert or translate a general functional behavior encoded by a coherent superposition of functional clauses
(and made injective by adding bits) into orthogonal subspaces of a Hilbert space has been found.
Also~the method of dimensional lifting introduced here is quantum infeasible because it is non-unitary (translating non-orthogonal vector into orthogonal ones).
Therefore, the question remains of whether a~general unitary method of dimensional lifting, possibly utilizing auxiliary bits, can be found.

\acknowledgments{\textbf{Acknowledgments:} This work was supported in part by the European Union, Research Executive Agency (REA),
Marie Curie FP7-PEOPLE-2010-IRSES-269151-RANPHYS grant. }


\ifws
\bibliographystyle{mdpi}
\fi


%

\end{document}